\documentstyle[12pt,aasms4,psfig]{article} 
\footnotesep \baselineskip      
\def\ein{{\it Einstein~}}
\def\ginga{{\it Ginga~}}
\def\rosat{{\it ROSAT~}}
\def\exosat{{\it EXOSAT~}}
\def\tenma{{\it Tenma~}}
\def\asca{{\it ASCA~}}

\def\la{\hbox{\rlap{$<$}\lower.5ex\hbox{$\sim$}\ }}

\received{24$^{\rm th}$ February 1996}
\accepted{5$^{\rm th}$  April 1996}

\slugcomment{To appear in the June 20$^{\rm th}$ issue of the Astrophysical Journal Letters.}

\lefthead{Harrus et al.}
\righthead{Discovery of an X-ray Synchrotron Nebula Associated with the Radio Pulsar PSR B1853+01 in the Supernova Remnant W44}

\begin{document}
\title{Discovery of an X-ray Synchrotron Nebula Associated with the Radio Pulsar PSR B1853+01 in the Supernova Remnant W44}

\author{Ilana~M. Harrus\altaffilmark{1},  John~P. Hughes\altaffilmark{2}}
\affil{ Harvard-Smithsonian Center for Astrophysics}
\affil{60 Garden Street, Cambridge, MA 02138}
\author{ and  David~J. Helfand}  
\affil{ Astronomy Department, Columbia University}
\affil{ 538 West 120$^{\rm th}$ Street, New York, NY 10027}
\altaffiltext{1}{Department of Physics, Columbia University, New York, NY 10027}
\altaffiltext{2}{E-mail: jph@cfa.harvard.edu} 
\begin{abstract}
We report the detection using data from the {\it Advanced Satellite for
Cosmology and Astrophysics} (\asca) of a hard X-ray source in the 
vicinity of the radio pulsar PSR B1853+01, which is located within the
supernova remnant (SNR) W44. PSR B1853+01, a 267~ms pulsar, has to date
been detected only in the radio band.  Previous observations at soft
X-ray energies (e.g., with  \rosat HRI) have failed to detect any
significant X-ray emission (pulsed or unpulsed) from the pulsar. In
addition, no high energy emission (${_>\atop^{\sim}}$4~keV) has been detected
previously from W44.

Over the 0.5--4.0~keV band, the \asca data show soft thermal
emission from W44, with a morphology very similar to
that observed earlier by \ein  and {\it ROSAT}.
In the high-energy band (4.0--9.5 keV), the SNR is, for the most part, 
invisible, although a source coincident with the position of PSR
B1853+01 is evident. The observed \asca spectra are consistent
with a power-law origin (photon index $\sim$2.3) for the X-ray
emission from this source at a flux level (flux
density $\sim$0.5 $\mu$Jy at 1 keV) consistent with previous upper 
limits. 
The maximum allowed size for the source is determined directly
from the \asca data ($<$5$^\prime$), while the minimum size is
derived from the nondetection of a point source in the
\rosat HRI data (${_>\atop^{\sim}}$30$^\prime$$^\prime$).  
Timing analysis of the hard X-ray source  failed to detect pulsations at 
the pulsar's period.
Based on these lines of evidence, we conclude that the new hard source
in W44 represents an X-ray synchrotron nebula associated with PSR
B1853+01, rather than the beamed output of the pulsar itself.  This
discovery adds W44 to the small group of previously known plerionic
SNRs. This nebula lies at the low end of, but is
consistent with, the correlation between X-ray luminosity and pulsar
spin-down energy loss found for such objects, lending further support to our
interpretation. 
\end{abstract}

\section{ Introduction }
The radio pulsar PSR B1853+01 was discovered by Wolszczan, Cordes, \& Dewey
(1991) inside the radio shell of the supernova
remnant W44.  The estimated distances to the pulsar and the supernova remnant
(SNR) are both on the order of 3~kpc, and the age deduced from the spin-down of
the pulsar (20,000~yrs) is compatible with the dynamical age of the
remnant inferred from X-ray observations (Harrus \& Hughes 1994; Harrus et
al. 1996).
Previous work has failed to detect X-ray emission from
PSR B1853+01 in any of the expected forms: pulsed or unpulsed emission
from the pulsar itself, or nonthermal emission from an associated
synchrotron nebula.  No high energy tail was seen in the \ginga
spectrum, leading to a 3\,$\sigma$ upper limit of
$3.6\times 10^{-12}$~ergs~cm$^{-2}$~s$^{-1}$ for the 2--10~keV flux
of a Crablike power-law component ($dN/dE \sim E^{-2.1}$) from W44
(Harrus \& Hughes 1994; Harrus et al. 1996).  
Although there was a report of a high-energy
component to the X-ray spectrum of W44 from \exosat data (Rho et al.\
1994), an earlier analysis (Jones, Smith, \& Angellini
1993) had already indicated that this component arose
from contamination of the data by particle background events and that,
in fact, the SNR was undetected above 5~keV. Moreover neither a point
source nor pulsed emission was seen at the pulsar position in the
\rosat PSPC observation in the soft X-ray band (Rho et al.\ 1994).  In
this Letter, we report on the discovery of the anticipated nonthermal
X-ray emission from the vicinity of PSR B1853+01 using spectral,
imaging, and timing observations from {\it the Advanced Satellite
for Cosmology and Astrophysics} ({\it ASCA}) (Tanaka, Inoue, \& Holt 1994).

\section{ Analysis }

\asca conducted two observations of W44. The first, carried out on 
1994 April 22, had a nominal pointing direction of 18$^{\rm h}$56$^{\rm
m}$24$^{\rm s}$, 1$^\circ$16$^\prime$58$^{\prime\prime}$ (J2000).  In
anticipation of a search for pulsed X-ray emission from PSR B1853+01
(period $\simeq$ 267ms), we rearranged Gas Imaging Spectrometer (GIS)
telemetry bits in order to record photon arrival times more accurately
at the expense of pulse-height (PH) rise-time information. This
modification increased the GIS time resolution to 15.625~ms (medium
telemetry rate mode) and 1.95~ms (high telemetry rate mode) from their
nominal values of 500 and 62.5~ms, respectively.  Data from the GIS
were rejected during times when the satellite was traversing regions
of low geomagnetic rigidity ($\leq$7) and when the Earth elevation
angle was $\leq$10$^\circ$. The data were also masked spatially to
remove the bright ring of background around the edge of the detector
as well as the calibration source events.  The Solid-State
Imaging Spectrometer (SIS) time resolution of 8\,s and 16\,s, in the 2-CCD
and 4-CCD modes used in the observation was clearly too low to detect
pulsed emission from PSR B1853+01.  The SIS screening criteria used in
extracting spectral information were very similar to the ones applied
to the GIS: minimum Earth elevation angle of 10$^\circ$ and a minimum
rigidity of 6.  Because of the possibility for contamination in the
SIS of fluorescence lines of oxygen from the Earth's atmosphere, data
selection was done on the bright-Earth angle, and only data above
40$^\circ$ (20$^\circ$) for the SIS0 (SIS1) were retained. Finally,
only events with CCD grades 0, 2, 3, or 4 were used in further
analysis.

For the second observation, carried out on 1994 September 10, the
nominal pointing direction was changed to 18$^{\rm h}$55$^{\rm
m}$55$^{\rm s}$, 1$^\circ$28$^\prime$18$^{\prime\prime}$ and the GIS
telemetry bit assignments were set to their nominal values.
Consequently, the pulsar region was not included in the field of view
(FOV) of the SIS detectors and the time resolution of the GIS data was
inadequate for the pulsar search study, although the GIS imaging data
are included in this study. The total exposure time on W44 was
approximately 39 ks (GIS) and 48 ks (SIS) summing the two pointings.

Below, we present spatial, spectral, and timing analysis of these data, which
reveal an X-ray synchrotron nebula associated with the pulsar.

\subsection{Spatial Analysis}
\vspace{-0.1cm}
We generated exposure-corrected, background-subtracted merged images of the 
GIS and SIS data in selected spectral bands. Background was determined 
from the weighted average of several nominally blank fields from high Galactic 
latitude observations with data selection criteria matched to those used for
the W44 data. Exposure maps were generated from the off-axis effective-area
calibrations, weighted by the appropriate observation time. 
Events from regions of the merged exposure map with less than 10\%  
of the maximum exposure were ignored. Merged images of the source data,
background, and exposure were smoothed with a Gaussian of $\sigma$=
30$^{\prime\prime}$ for the low-energy band (0.5--4.0 keV) and $\sigma$ = 
90$^{\prime\prime}$ for the high-energy band (4.0--9.5 keV).
We subtracted smoothed background maps from the data maps and divided by the
corresponding exposure map. Figure~1 shows the results obtained for the two detectors in each of the two
energy bands. The low-energy band shows the soft thermal X-ray emission from W44,
with a morphology very similar to that observed before by \ein (Smith
et al.\ 1985) and \rosat (Rho et al.\
1994).  In the high-energy band, however, the SNR is not
visible and only a few unresolved sources appear. The most significant
of these is coincident with the position of the radio pulsar. In the
4.0--9.5 keV band, this source is detected with a signal-to-noise ratio
of $\sim$9, and at higher energies (6.0--9.5 keV), the signal-to-noise
ratio is $\sim$4. The other significant source, located about
10$^\prime$ northwest of the pulsar in the GIS image, has a
signal-to-noise ratio of $\sim$5.0 (4.0--6.0 keV); it is not
detected above 6 keV. We suggest that this second source arises from
temperature variation within the SNR itself and that we are seeing the
tail of the emission from a higher temperature parcel of shocked
plasma. (Note that the apparent difference in relative intensities of
these sources as observed by the SIS and GIS is most likely a result
of the different spectral resolution of the two detectors.  The SIS,
because of its better spectral resolution, can discriminate better
against out-of-band photons, e.g., those below 4 keV, than can the
GIS.) We will be addressing temperature variations in W44 in greater
detail in future work.
\clearpage
\setcounter{page}{8}
In the GIS (considering the sum of the GIS2 and GIS3 detectors), the
number of photons at energies greater than 4~keV within a 5$^\prime$
radius centered on the pulsar position is 310$\pm$21 counts (for an
exposure time 12,670~s) after subtracting a background image
derived from high Galactic latitude observations.  Elsewhere in the
FOV of the GIS outside the remnant boundaries, a similar size region
at a similar off-axis angle (9$^\prime$ for the test region compared
to 6$^\prime$ for the pulsar region) in the same band yields roughly
98$\pm$13 counts.  This excess of events in an apparently source-free
portion of the image is due to the Galactic plane X-ray background
emission, as we show in our spectral analysis below.  In the SIS, the
region selected for the pulsar analysis is smaller (radius of
3$^\prime$) in order not to cross individual chip boundaries. The
number of events seen at high energy (again summing the two detectors
SIS0 and SIS1) is 150$\pm$14, compared with 37$\pm$8 found in a  
source-free region elsewhere in the FOV of the SIS.
In Table~1, we have summarized 
the positions of the radio pulsar (Wolszczan 1995) and the
associated X-ray source, as measured by the two \asca detectors.  The
errors given in the table include both statistical and systematic
uncertainties.  The pulsar's position is indicated by a cross in
Figure~1.

\scriptsize
\centerline{\bf TABLE I} 
\vspace{-0.2cm}
\centerline{\bf    POSITION OF PSR B1853+01}
\vspace{-0.2cm}
\centerline{\bf  AND ASSOCIATED X-RAY SOURCE}
\vspace{0.1cm}
\centerline{\bf
\begin{tabular}{ccc} \hline\hline  
    & R.A.(J2000)  & Decl.(J2000) \\\hline  
Radio$^{\rm a}$ & 18${\rm ^h}$56${\rm ^m}$10.79${\rm ^s}$$\pm$0.04${\rm ^s}$ &
1$^\circ$13$^\prime$28$^{\prime\prime}$$\pm$2$^{\prime\prime}$ \\  
X-ray:GIS  & 18${\rm ^h}$56${\rm ^m}$10.0${\rm ^s}$~$\pm$1.5${\rm ^s}$ &
1$^\circ$13$^\prime$17$^{\prime\prime}$~$\pm$40$^{\prime\prime}$ \\ 
X-ray:SIS & 18${\rm ^h}$56${\rm ^m}$12.6${\rm ^s}$~$\pm$1${\rm ^s}$
&1$^\circ$12$^\prime$34$^{\prime\prime}$~$\pm$30$^{\prime\prime}$\\ \hline 
\end{tabular}} 
\vspace{0.1cm} 
\centerline{$^{\rm a}$  \bf Wolszczan 1995}
\normalsize 
A recent high-resolution radio map of W44 shows an elongated diffuse
feature with a flat spectral index ($-$0.12$\pm$0.04) extending north
from the pulsar's position (Frail et al.\ 1996). The
peak of this emission is centered at approximately 18$^{\rm
h}$56$^{\rm m}$11$^{\rm s}$, 1$^\circ$13$^\prime$24$^{\prime\prime}$
and it is extended over an elliptical region $\sim$1$^\prime$ in right
ascension and $\sim$2$^\prime$ in declination.  We note that the
position of the new hard X-ray source is consistent with that of the
pulsar, while it is marginally inconsistent with the peak position of
the diffuse radio nebula to the north.
\subsection{Spectral analysis}

Our previous work on W44 (Harrus \& Hughes 1994; Harrus et al. 1996) 
has shown that the
remnant's global integrated X-ray emission can be described adequately
using a simple nonequilibrium ionization plasma model 
(Hughes \& Singh 1994),
 characterized by a temperature $kT = (0.88 \pm 0.14)$~keV, an
ionization timescale $n_et$ =
(2.0$^{+4.3}_{-0.7}$)$\times$10$^{11}$~cm$^{-3}$~s, and a column
density $N_{\rm H}$~=~(1.0$^{+0.6}_{-0.2}$)$\times$10$^{22}$~atoms~cm$^{-2}$.
The errors are quoted at the 90\% confidence
level.  In this model, the ionization timescale characterizes
deviations from ionization equilibrium; the larger the value, the
closer the plasma is to equilibrium, which occurs at a value
$\sim$3$\times$10$^{12}$~cm$^{-3}$~s. For simplicity, and in view of
the large value of the ionization timescale, we have approximated the
SNR spectrum with a single-temperature Raymond \& Smith (1977;
version 9.00 in XSPEC) thermal plasma model. Line-of-sight absorption
by the interstellar medium is included using the cross sections from
Morrison \& McCammon (1983). 

The Galactic plane is a strong source of diffuse  X-ray emission that
can be modeled well in the 2--10 keV band by a thermal bremsstrahlung
continuum with a temperature $kT \sim 7$ keV, and a narrow emission line
from highly ionized iron ($E\sim 6.8$ keV), with an equivalent width
$\sim$ 0.6 keV (Yamauchi 1991).  Previous observations of roughly this
area of the sky by \tenma\ (Koyama 1989) and \ginga\ 
(Yamauchi 1991) suggest
a surface brightness in the range 
(0.56--2.26)~$\times$10$^{-7}$~ergs~cm$^{-2}$~s$^{-1}$~sr$^{-1}$ 
for the plane emission (2--10 keV band). These
instruments, however, had large fields of view ($\sim$3$^\circ$ FWHM
for {\it Tenma}, and $1^\circ \times 2^\circ$ FWHM for {\it Ginga}), and it is
not known how structured the plane emission is on smaller angular
scales ($\sim$5$^\prime$). In order to normalize the intensity of the
plane emission for the \asca analysis, we selected a ``control''
region of the same size as the pulsar region, located
elsewhere in the field of view, making sure that the control and pulsar
regions were disjoint.  Our procedure entailed determining the intensity of
the Galactic plane emission (for a fixed set of spectral parameters)
from the control region, and then applying that normalization and set of 
spectral parameters to the X-ray spectrum extracted from the pulsar region.

Figure~2 shows the results of the fit and the residuals associated with
these models. In this analysis, as well as in the timing analysis
discussed below, we have used data from the first observation taken in
medium-bit-rate mode.  Using the intensity of the Galactic plane
emission deduced from the fit of the control region, we study the data
extracted from a circular region of radius 5$^\prime$ (3$^\prime$) in
the GIS (SIS), centered on the radio pulsar position. First, we fit the
data in the pulsar region with a Raymond \& Smith (1977) plasma model, plus
the Galactic plane background model, and find a $\chi^2$ of 622 (567) and a 
reduced $\chi^2$ of 1.8 (4.0), for the GIS (SIS) data.  Then we fix
the parameters at their best-fit values and include an additional
power-law component fitted to the band above 3~keV only.  The
$\Delta\chi^2$ (computed over the whole range of energy) of these
fits is greater than 30 for the GIS and greater than 50 for the SIS (a
reduction significant at more than the 99\% confidence level); the
power-law index is 2.3$^{+1.1}_{-0.9}$, and its normalization is
0.7$^{+2.3}_{-0.5}$$\times$10$^{-3}$~photons~cm$^{-2}$~s$^{-1}$~keV$^{-1}$
at 1~keV in the GIS
(0.6$^{+2.1}_{-0.5}$$\times$10$^{-3}$~photons~cm$^{-2}$~s$^{-1}$~keV$^{-1}$
at 1~keV in the SIS).  The errors on the slope of the power-law combine
the statistical error (which dominates) and the uncertainties in the
Galactic background model used.  We varied the temperature of the
background bremsstrahlung continuum between 6 and 8~keV, and we also varied
the normalization of the model at a fixed temperature. The
uncertainties are then combined in quadrature.  The associated remnant
temperature is 0.5$\pm$0.2~keV, and the column density is
(1.82$\pm$0.05)$\times$10$^{22}$~atoms~cm$^{-2}$.  We show the results
for the GIS only in Figure~3; GIS 2 and 3 have been merged for display
purposes only.

\clearpage
\setcounter{page}{12}
\subsection{Timing analysis} 

We have carried out a timing analysis on events recorded in the two
GIS with a time resolution of 15.6~ms. We select only the 453 events
coming from the same region as the one used in the spectral analysis
and having an energy larger than 4~keV (to discriminate against
photons coming from the SNR itself).  Folding the data with the known
radio timing parameters of {P} = 0.26743520599(6)~s and \.{P} of
(208.482$\pm$~0.006)$\times$10$^{-15}$ s~s$^{-1}$ (Wolszczan 1995), we
find no significant modulation at the 3\,$\sigma$ level.

One limitation of such a folding method is its dependence on bin
size: the data are binned according to the phase difference between
the corrected arrival time and the period. Considering the small
number of photons available, we have also performed a Z$_n$ test
(Buccheri et al.\ 1983) summing harmonics with
 $n\le4$.  The Z$_n$ test is free of the bias of the bin size chosen in
the folding method and is sensitive to a range of pulse shapes
depending on the order of the test (larger $n$ implies sensitivity to
narrower pulse shapes).  We have used simulated data to put an upper
limit of 10\% on the pulsed fraction for the emission above 4~keV.
This limit depends only weakly on the order of the Z$_n$ test
performed.

\section{Discussions and Conclusions} 

We have presented the results of morphological and spectral studies of
the contribution from PSR B1853+01 to the X-ray emission of the SNR
W44.  High-energy emission in both the GIS and the SIS is found at the
position of the pulsar; the spectra from both instruments require a
power-law component to model the observed emission.  A lower limit to
the spatial extent of the emission region can be derived from the
average \rosat HRI surface brightness around the pulsar position, 
which, after subtraction of the nominal HRI background, is
 $\sim$4$\times$10$^{-3}$ counts~arcmin$^{-2}$~s$^{-1}$.  Based on our
spectral analysis, we expect an HRI count rate from the power-law
component of (2.0--3.5)$\times$10$^{-3}$ counts~s$^{-1}$.  This
translates into a minimum angular size of $\sim$30$^{\prime\prime}$.
Using the known distance to the SNR, we deduce an emission volume of
$2.78\,D_{\rm 3\,kpc}^3\theta^3\ {\rm pc}^3$, where $D_{\rm 3\,kpc}$
is the distance to the SNR in units of 3~kpc, and $\theta$ is the
angular size of the nebula expressed in arcminutes.  An upper limit on the
source size can be obtained from the \asca data.  We have simulated
images with Gaussian source profiles of various widths and convolved
each with the PSF of the mirror and the detector, weighted according
to the spectrum from the pulsar region.  We then cross-correlated the
simulated images with the GIS image and deduced an upper limit to the
source radius of $\sim$5$^\prime$.

The inferred unabsorbed flux from the power-law component extrapolated
to the \rosat energy band (0.4--2.0~keV) is
1.9$_{-1.4}^{+9.1}$$\times$10$^{-12}$ ergs~cm$^{-2}$~s$^{-1}$; the
value of 1.2$\pm$0.3$\times$10$^{-12}$ ergs~cm$^{-2}$~s$^{-1}$, 
measured for the 2--10 keV band flux, is in agreement with the upper limit
of 3.6$\times$10$^{-12}$ ergs~cm$^{-2}$~s$^{-1}$ obtained using the
\ginga spectrum (Harrus \& Hughes 1994; Harrus et al. 1996). The
unabsorbed X-ray luminosity in the \ein band (0.2--4.0~keV) is
(4$_{-3}^{+30}$)$\times10^{33}\,\,D_{\rm 3\,kpc}^2$~ergs~s$^{-1}$, 
which should be compared to the luminosity predicted by the empirical
\.{E}/L$_{X}$ relation (Seward \& Wang 1988) of 7.7$\times$10$^{32}$
ergs~s$^{-1}$.

Frail et al.\ (1996) present new radio data on the nebula surrounding
PSR B1853+01 and, using the \ginga\ upper limit, construct a spectrum
covering 10 decades in frequency from the radio to the X-ray bands.
From these data, the equipartition value of the nebular magnetic
field, ${\rm B}_n$, and the energy in relativistic electrons, $E_e$,
can be estimated by assuming that the break in the spectrum between
the radio and X-ray regimes is due to synchrotron losses 
(Pacholczyk 1970, p~169).
Our \asca measurements of the X-ray flux and spectral index of
the nebula allow better determination of the break frequency, that is,
the frequency where the extrapolated X-ray and radio power-law spectra
intersect. The best-fit spectral values indicate a value of $\nu_{\rm
B} \sim 2 \times 10^{13}$ Hz. (Note that the break frequency of the
Crab Nebula is $\sim$$10^{13}$ Hz.)  We find for the nebular magnetic
field ${\rm B}_n \simeq 70\,\mu{\rm G}\, (V_{\rm R} /
3.1\times10^{55}\,{\rm cm}^3)^{-2/7}$ (using the estimated volume of
the radio nebula, $V_{\rm R}$, and the luminosity integrated from
$10^7$ Hz to $\nu_{\rm B}$ which is $2.3\times 10^{34}$ ergs s$^{-1}$)
and for the energy in electrons 
 $E_e \simeq 10^{46}\,{\rm ergs}\, ({\rm B}_n / 70\,\mu{\rm G})^{-3/2}$, values that are consistent with other plerionic SNRs of
comparable luminosity.

The lifetime of the electrons giving rise to the radio emission,
$\sim$20,000 yr, is of the same order as the age of the SNR and
pulsar, while the electrons giving rise to the X-ray emission are
quite short-lived, $\sim$120 yr.  Consequently, the X-ray synchrotron
nebula should be significantly smaller than the radio nebula and
should be located close to the pulsar, as indicated by our \asca\
data. In particular, given the parameters derived above, we expect
essentially no nonthermal X-ray emission from the bright diffuse
radio-emitting nebula north of the pulsar, since the pulsar would have
traversed this region over 2000 years ago, given the projected distance
and the inferred transverse motion, 375 km s$^{-1}$, of PSR B1853+01
(Frail et al.\ 1996).  A follow-up X-ray observation of the pulsar
region with high spatial resolution should be able to confirm this
conjecture and, in addition, provide a number of new constraints on
the energetics of the synchrotron nebula.  In any event, the \asca\
data presented here provide strong evidence for the oldest synchrotron
nebula yet detected, a result that should aid considerably in our
attempts to understand the interaction of pulsar relativistic winds
with their environments.

We thank Alex Wolszczan for providing an accurate position and a
recent ephemeris for PSR B1853+01 and we thank E.~Churazov, 
M.~Gilfanov and A.~Finoguenov for allowing us to use their software for
merging \asca images.  Pat Slane, Olaf Vancura, Paul Callanan, Didier
Barret, Rick Harnden, and Fred Seward are acknowledged for their
helpful comments and discussions.  This research was partially
supported by NASA Grant NAG 5-2605 and the Smithsonian Institution, 
through the Smithsonian Predoctoral Fellowship program. This is
contribution 595 of the Columbia Astrophysics Laboratory.


\end{document}